\documentclass[12pt]{iopart}

\usepackage{iopams}
\usepackage{graphicx}
\usepackage{color}
\usepackage{mathrsfs}
\usepackage{bm}

\begin{document}

\title{On the gyrocenter transform implemented on Lagrangian differential 1-form with the existence of magnetic perturbation}

\author{Shuangxi Zhang }

\address{
 Graduate School of Energy Science, Kyoto University, Uji, Kyoto 611-0011, Japan. }
\ead{zhang.shuangxi.3s@kyoto-u.ac.jp; zshuangxi@gmail.com}

\date{\today}

\begin{abstract}
This paper pointed out that the usual gyrocenter transform, which implements single-parameter Lie transform perturbation theory on Lagrangian differential 1-form of the orbit of charged particles immersed in a strong magnetic field, is not a near identical coordinate transform, and further causes some unphysical terms in the eventual orbit equations.
\end{abstract}


\maketitle

\section{Introduction}\label{sec1}

For decades gyrokinetic theory is treated  by the theoretical and simulation community of magnetically confined fusion plasmas as a strong tool to simplify the physics model of magnetized plasmas by reducing the gyrophase of each particle\cite{1982frieman,1983wwlee,1983dubin,wwleejcp1987,1988hahm,1990brizard,2000sugama,2000qinhong1,1983littlejohn}. This simplification hugely reduces the computing burden of numerical simulation\cite{wwleejcp1987,1998linzhihong,2000jenko,2008idomura,2009peeters,2010garbet}. To reduce the gyrophase of each particle from the whole dynamical system, the single-parameter Lie transform perturbation theory (SPLTPT) is implemented\cite{1983cary,1988hahm,1990brizard}. Gyrokinetic theory in the past literature involves two independent but consecutive coordinate transforms\cite{1990brizard,2010scott}. The first one is guiding center transform; the second one is the gyrocenter transform. There are two approaches involved in the application
of SPLTPT for the gyrocenter transform: one is applying SPLTPT to the Hamiltonian on guiding-center coordinate\cite{1981cary,1983dubin,1988hahm2}; the other one involves applying SPLTPT to the Lagrangian 1-form\cite{1990brizard,2000qinhong1}, which determines the orbit equation of charged particles on guiding-center coordinate. Both approaches are widely applied. The second approach is the focus of this paper, and it will be pointed out that for magnetic perturbation, the coordinate transform derived from this approach is not a near identical transform (NIT), which further induces some unphysical terms. The analysis of the first approach will be given in another paper. For the second approach, the transform of the Lagrangian 1-form is realized by carrying out the pullback transform over the Lagrangian 1-form, with the ultimate goal of reducing the gyrophase from the whole dynamical system.

The rest of this paper is arranged as follows. In Sec.(\ref{sec2}), SPLTPT is carried out on the normalized Lagrangian differential 1-form and the orbit equations are obtained.  Sec.(\ref{sec3}) points out the violation of NIT by this coordinate transform and the unphysical terms as the consequence. The appendix gives a simple introduction of SPLPTP.

\section{Carry out the pullback transform over the Lagrangian 1-form}\label{sec2}


\subsection{Normalizing physics quantities}\label{sec2.1}

The basic formula which will be used is the Lagrangian differential 1-form of the motion of charged particle chosen from the magnetized plasmas
\begin{equation}\label{a119}
\gamma ' = \left( {q{\bf{A}}\left( {{{\bf{x}}}} \right) + m{\bf{v}}} \right)\cdot d{\bf{x}} - \frac{1}{2}m{v^2}dt.
\end{equation}
$(\mathbf{x},\mathbf{v})$ is the full particle coordinate frame.
By decoupling the gyroangle $\theta$ from other degrees of freedom up to $O(\varepsilon^2)$ with $\varepsilon  = \frac{{\rho} }{{{L_0}}}$ and $\rho$ being the Larmor radius, it gives the guiding center Lagrangian 1-form like
\begin{eqnarray}\label{g88}
{\gamma _0} = && \left( {q{\bf{A}}\left( {{{\bf{X}}_{\bf{1}}}} \right) + m{U_1}{\bf{b}}} \right)\cdot d{{\bf{X}}_1} + \frac{m}{q}{\mu _1}d{\theta _1} \nonumber \\
&&- ({\mu _1}B\left( {{{{\bf{X}}}_{{1}}}} \right) + \frac{1}{2}m U_1^2)dt,
\end{eqnarray}
where $\mathbf{A}(\mathbf{X}_1)$ is the equilibrium magnetic potential. Then, differential 1-form for the perturbative wave is introduced
\begin{equation}\label{g89}
{\gamma _w} = q{{\bf{A}}_w}({{\bf{X}}_1} + {\bm{\rho }},t)\cdot d\left( {{{\bf{X}}_1} + {\bm{\rho }}} \right) - q\phi_w \left( {{{\bf{X}}_1} + {\bm{\rho }},t} \right)dt,
\end{equation}
with ${\bm{\rho }} = {{\bm{\rho }}_0} + ( \cdots )$ and ${\bm{\rho} _0} = \frac{1}{q}\sqrt {\frac{{2m{\mu _1}}}{{B\left( {{{\bf{X}}_1}} \right)}}} \left( { - {{\bf{e}}_1}\cos \theta_1  + {{\bf{e}}_2}\sin \theta_1 } \right)$. $(\mathbf{b},\mathbf{e}_1,\mathbf{e}_2)$ forms a right-hand cartesian coordinate frame. $\mathbf{b}$ is the unit vector directing along equilibrium magnetic field $\mathbf{B}(\mathbf{X}_1)$, and $\theta_1$ is the gyroangle. The rotation direction of ions around the magnetic field line is inverse to that of electrons.
Symbol $``(\cdots)"$ means higher order terms. $\mathbf{A}_w,\phi_w$ denote the perturbations of the magnetic potential and the electric potential, respectively. Here, $\mathbf{A}(\mathbf{X_1})$ is the equilibrium magnetic potential, and  the guiding center coordinates plus the time is denoted as ${{\bf{Z}}}_1 \equiv ({{\bf{X}}}_1,{U}_1,{\mu}_1 ,{\theta}_1, t)$. The other notations in Eqs.(\ref{g88},\ref{g89}) can be referred in Ref.(\cite{1990brizard}).

The test particle is chosen from a thermal equilibrium plasma ensemble, e.g., the thermal equilibrium plasma in tokamak. Therefore, $\mathbf{A},U_1,\mathbf{X}_1,t,\mathbf{B}, \phi_w, \mu_1$ can be normalized by $A_0=B_0 L_0,v_t,L_0,L_0/v_t,B_0,A_0 v_t, mv_t^2/B_0$, respectively. $B_0, L_0$ are characteristic amplitude and spatial length of the magnetic field, respectively. $v_t$ is the thermal velocity of the particle ensemble which contains the test particle. The small parameter representing the normalized amplitude of $\mathbf{A}_w,\phi_w$ is extracted out, so that $\mathbf{A}_w,\phi_w$ are reformulated as $\varepsilon_w \mathbf{A}_w, \varepsilon_w \phi_w$, respectively, with $O(|\mathbf{A}_w|)\sim O(|\phi_w|)\sim O(1)$.  Here, it's assumed that $\mathbf{A}_w$ and $\phi_w$ are of the same amplitude. Throughout the rest of the paper, all physical quantities are normalized.

The detailed normalization procedure is given by taking Eq.(\ref{g88}) as an example. First, divide both sides of Eq.(\ref{g88}) by $m{v_t}{L_0}$. The first term of RHS of Eq.(\ref{g88}) is like $\frac{{q{A_0}}}{{m{v_t}}}\frac{{{\bf{A}}\left( {{{\bf{X}}_1}} \right)}}{{{A_0}}}\cdot\frac{{d{{\bf{X}}_1}}}{{{L_0}}}$, which is further written as $\frac{1}{\varepsilon }{\bf{A}}\left( {{{\bf{X}}_1}} \right)\cdot d{{\bf{X}}_1}$, with the replacement: $\varepsilon  \equiv \frac{{m{v_t}}}{{q{A_0}}}$, $\frac{{{\bf{A}}\left( {{{\bf{X}}_1}} \right)}}{{{A_0}}} \to {\bf{A}}\left( {{{\bf{X}}_1}} \right),\frac{{d{{\bf{X}}_1}}}{{{L_0}}} \to d{{\bf{X}}_1}$. Other terms can be normalized in the same way. Eventually, we could derive a normalized Lagrangian 1-form like
\begin{eqnarray}
\frac{{{\gamma _0}}}{{m{v_t}{L_0}}} =& \left( {\frac{1}{\varepsilon}{\bf{A}}\left( {{{\bf{X}}_1}} \right) + {U_1}{\bf{b}}} \right)\cdot d{{\bf{X}}_1} + \varepsilon{\mu _1}d{\theta _1} \nonumber \\
& - ({\mu _1}B\left( {{{\bf{X}}_1}} \right) + \frac{1}{2}U_1^2)dt,
\end{eqnarray}
by utilizing the normalization scheme given previously. Now, multiplying both sides by $\varepsilon$, and rewriting $\frac{{{\varepsilon \gamma _0}}}{{m{v_t}{L_0}}}$ to be $\gamma_0$, we derive the normalized 1-form as follows
\begin{eqnarray}\label{g2}
{\gamma _0} =&& {\bf{A}}\left( {{{\bf{X}}_1}} \right)\cdot d{{\bf{X}}_1} + \varepsilon {U_1}{\bf{b}}\cdot d{{\bf{X}}_1} + {\varepsilon ^2}{\mu _1}d{\theta _1} \nonumber \\
&& - \varepsilon \left( {\frac{{U_1^2}}{2} + {\mu _1}B\left( {{{\bf{X}}_1}} \right)} \right)dt.
\end{eqnarray}
Since a constant factor $\frac{\varepsilon }{{m{v_t}{L_0}}}$ doesn't change the dynamics determined by the Lagrangian 1-form, the Lagrangian 1-form given by Eq.(\ref{g2}) is of the same dynamics with that given by Eq.(\ref{g88}).
By utilizing the same normalization procedure, with $\mathbf{A}_w,\phi_w$ changed to be $\varepsilon_w \mathbf{A}_w, \varepsilon_w\phi_w$, respectively, Eq.(\ref{g89}) becomes
\begin{eqnarray}\label{g3}
&&{\gamma _w} = \varepsilon_w {{\bf{A}}_w}\left( {{{\bf{X}}_1} + \bm{\rho} ,t} \right)\cdot d\left( {{{\bf{X}}_{\bf{1}}} + \bm{\rho} } \right) + \varepsilon_w {\phi _w}\left( {{{\bf{X}}_{\bf{1}}} + \bm{\rho} ,t} \right)dt \nonumber \\
&& \approx \varepsilon_w \exp \left( {{\varepsilon \bm{\rho} _0}\cdot\nabla } \right){{\bf{A}}_w}\left( {{{\bf{X}}_1},t} \right)\cdot\left( \begin{array}{l}
d{{\bf{X}}_1} + \frac{{\varepsilon\partial { \bm{\rho} _0}}}{{\partial {{\bf{X}}_1}}}\cdot d{{\bf{X}}_1} \nonumber \\
 + \frac{{\varepsilon \partial {{\bm{\rho}} _0}}}{{\partial {\mu _1}}}d{\mu _1} + \frac{{\varepsilon \partial { \bm{\rho} _0}}}{{\partial {\theta _1}}}d{\theta _1}
\end{array} \right)  \nonumber \\
&&  - \varepsilon_w \exp \left( {{\varepsilon \bm{\rho} _0}\cdot\nabla } \right){\phi _w}\left( {{{\bf{X}}_1},t} \right)dt,
\end{eqnarray}
where ${\varepsilon } \equiv \frac{{m{v_t}}}{{{A_0}q}}=\frac{{{\bm{\rho} _t}}}{{{L_0}}}$, ${\bm{\rho} _t} = \frac{{m{v_t}}}{{{B_0}q}}$, ${\bm{\rho} _0} = \sqrt {\frac{{2\mu }}{{B\left( {\bf{X}}_1 \right)}}} \left( { - {{\bf{e}}_1}\cos \theta  + {{\bf{e}}_2}\sin \theta } \right)$. In Eqs.(\ref{g2},\ref{g3}), all $\varepsilon,\varepsilon_w$ takes part in calculation. If the small parameters $\varepsilon,\varepsilon_w$ are just used as a symbol of the order of terms, they are denoted as $\varepsilon^*, \varepsilon_w^*$. This rule is adopted throughout the rest of this paper.

\subsection{Carrying out the pullback transform and deriving the  orbit equations of the gyrocenter}\label{sec2.2}

$\gamma_{0}+\gamma_{w}$ is the total Lagrangian differential 1-form with $\rho_0$ depending on the fast angle $\theta$.  To reduce $\theta$ from the whole coordinate system, SPLTPT given in \ref{app1} is adopted with $\varepsilon_w$ treated as the small parameter, while $\varepsilon$ as a normal quantity not involved in the order expanding. The gyrocenter frame is recorded as ${\bf{Z}} = \left( {{\bf{X}},\mu ,U,\theta } \right)$.
The coordinate transform should satisfy NIT and is formally recorded  as ${{\bf{Z}}_1} = \exp \left( { - {\varepsilon _w}{g^i}\left( {\bf{Z}} \right){\partial _i}} \right){\bf{Z}}$ with  $O(g^i)\sim O(1)$ for all $i\in \{\mathbf{X},U,\mu,\theta\}$. All of $g^i$s need to be solved. The new $\Gamma$ induced by this coordinate transform is
\begin{equation}\label{a99}
\Gamma=[\cdots T_2 T_1(\gamma_0+\gamma_w)](\mathbf{Z})+dS,
\end{equation}
with ${T_i} = \exp \left( { - \varepsilon _w^j{L_{{{\bf{g}}_j}}}} \right)$. $\mathbf{g}_j$ includes elements $g_j^i$ for $i\in \{\mathbf{X},\mu,U,\theta\}$.
By expanding $\Gamma$ in Eq.(\ref{a99}) as the sum like
\begin{equation}
\Gamma  = \sum\limits_{n\ge 0} {\frac{1}{{n!}}\varepsilon _w^n{\Gamma _n}},
\end{equation}
$O(\varepsilon_w^0)$ part of the new $\Gamma$ is
\begin{equation}\label{g35}
{\Gamma _0} = {{\bf{A}}}\left( {\bf{X}} \right)\cdot d{\bf{X}} + {\varepsilon }U{\bf{b}}\cdot d{\bf{X}} + {\varepsilon ^{2}}\mu d\theta
 - {\varepsilon }H_0 dt,
\end{equation}
with $H_0={\frac{{U_{}^2}}{2} + \mu B\left( {\bf{X}} \right)}$. The $O(\varepsilon_w)$ part is
\begin{eqnarray}\label{g36}
{\varepsilon _w}{\Gamma _1} =&& \left( { - \left( {{\bf{B}} + \varepsilon U\nabla  \times {\bf{b}}} \right) \times \left( {{\varepsilon _w}{{\bf{g}}^X}} \right) - \varepsilon {\varepsilon _w}{g^U}{\bf{b}} + \exp \left( {{\varepsilon \bm{\rho} _0}\cdot{\nabla _{\bf{X}}}} \right)\left( {{\varepsilon _w}{{\bf{A}}_w}} \right)} \right)\cdot d{\bf{X}} \nonumber \\
&& + \varepsilon \left( {{\varepsilon _w}{{\bf{g}}^X}\cdot{\bf{b}}} \right)dU + \left( {\exp \left( {{\varepsilon \bm{\rho} _0}\cdot{\nabla _{\bf{X}}}} \right)\left( {{\varepsilon _w}{{\bf{A}}_w}} \right)\cdot\frac{{\partial {\varepsilon \bm{\rho} _0}}}{{\partial \theta }} - {\varepsilon ^2}{g^\mu }} \right)d\theta  \nonumber \\
&& - \left( {{\varepsilon ^2}\left( {{\varepsilon _w}{g^\theta }} \right) + \exp \left( {\varepsilon {\bm{\rho} _0}\cdot{\nabla _{\bf{X}}}} \right)\left( {{\varepsilon _w}{{\bf{A}}_w}} \right)\cdot\frac{{\varepsilon\partial { \bm{\rho} _0}}}{{\partial \mu }}} \right)d\mu  \nonumber \\
&& - \left( { - \left( {{\varepsilon _w}{{\bf{g}}^X}\cdot\nabla {H_0}} \right) - \varepsilon {\varepsilon _w}U{g^U} - \varepsilon {\varepsilon _w}{g^\mu }B + \exp \left( {{\varepsilon \bm{\rho} _0}\cdot{\nabla _{\bf{X}}}} \right)\left( {{\varepsilon _w}{\phi _w}} \right)} \right)dt  \nonumber \\
&& + {\varepsilon _w}dS.
\end{eqnarray}
Eq.(\ref{g36}) obviously shows the confusion between the order of $\varepsilon$ and $\varepsilon_w$.

Modern GT requires the following identities
\begin{equation}\label{g34}
\Gamma_{1i}=0,i\in \{\mathbf{X},U,\mu,\theta\}.
\end{equation}
The $\mathbf{X}$ component of generators can be derived based on Eqs.(\ref{g36},\ref{g34}) as
\begin{eqnarray}\label{g37}
{{\bf{g}}^X} = && - \frac{1}{{{\bf{b}}\cdot{{\bf{B}}^*}}}\left( {{\bf{b}} \times \exp \left( {\varepsilon \bm{\rho}_0 \cdot\nabla } \right){{\bf{A}}_w}\left( {{\bf{X}},t} \right)  + {\bf{b}} \times \nabla {S_1}} \right)  \nonumber \\
&&- \frac{{{{\bf{B}}^*}}}{{{\varepsilon }}}\frac{{\partial {S_1}}}{{\partial U}}
\end{eqnarray}
where
\begin{equation}\label{f2}
{{\bf{B}}^*} = {\bf{B}} + {\varepsilon }U\nabla  \times {\bf{b}}.
\end{equation}
Other generators are given below
\begin{equation}\label{g38}
{g^U} = \frac{1}{{{\varepsilon }}}{\bf{b}}\cdot\exp \left( {\varepsilon \bm{\rho}_0 \cdot\nabla } \right){{\bf{A}}_w}\left( {{\bf{X}},t} \right) + \frac{1}{{{\varepsilon }}}{\bf{b}}\cdot\nabla {S_1},
\end{equation}
\begin{equation}\label{g39}
{g^\mu } = \frac{1}{{{\varepsilon}}}\exp \left( {\varepsilon \bm{\rho}_0 \cdot\nabla } \right){{\bf{A}}_w}\left( {{\bf{X}},t} \right)\cdot\frac{{ \partial \bm{\rho}_0 }}{{\partial \theta }} + \frac{1}{{{\varepsilon ^{2}}}}\frac{{\partial {S_1}}}{{\partial \theta }},
\end{equation}
\begin{equation}\label{g40}
{g^\theta } =  - \frac{1}{{{\varepsilon}}}\exp \left( {\varepsilon \bm{\rho}_0 \cdot\nabla } \right){{\bf{A}}_w}\left( {{\bf{X}},t} \right)\cdot\frac{{\partial \bm{\rho}_0 }}{{\partial \mu }} - \frac{1}{{{\varepsilon ^{2}}}}\frac{{\partial {S_1}}}{{\partial \mu }}.
\end{equation}
The equation for the gauge function is
\begin{eqnarray}\label{g41}
\frac{{\partial {S_1}}}{{\partial t}} + U{\bf{b}}\cdot\nabla {S_1} + \frac{1}{{{\varepsilon }}}\frac{{\partial {S_1}}}{{\partial \theta }} = F + {\Gamma _{1t}},
\end{eqnarray}
where
\begin{eqnarray}\label{g92}
F = && \exp \left( {\varepsilon \bm{\rho}_0 \cdot\nabla } \right){\phi _w}\left( {{\bf{X}},t} \right) \nonumber \\
&& -  U{\bf{b}}\cdot\exp \left( {\varepsilon \bm{\rho}_0 \cdot\nabla } \right){{\bf{A}}_w}\left( {{\bf{X}},t} \right) \nonumber \\
&&- {B(\mathbf{X})}\exp \left( {\varepsilon \bm{\rho}_0 \cdot\nabla } \right){{\bf{A}}_w}\left( {{\bf{X}},t} \right)\cdot\frac{{\partial \bm{\rho}_0 }}{{\partial \theta }}.
\end{eqnarray}
The smaller term ${{{\bf{g}}^X}\cdot \varepsilon \nabla {H_0}}$ and other higher order terms are ignored in Eq.(\ref{g41}).

For the low frequency perturbation, inequalities $\left| {\frac{{\partial {S_1}}}{{\partial t}}} \right| \ll \left| {\frac{B}{{{\varepsilon }}}\frac{{\partial {S_1}}}{{\partial \theta }}} \right|, \left| {U{\bf{b}}\cdot\nabla {S_1}} \right| \ll \left| {\frac{B}{{{\varepsilon }}}\frac{{\partial {S_1}}}{{\partial \theta }}} \right|$ hold, and the lowest order equation of Eq.(\ref{g41}) is
\begin{eqnarray}\label{g42}
\frac{B(\mathbf{X})}{{{\varepsilon }}}\frac{{\partial {S_1}}}{{\partial \theta }} = F +\Gamma_{1t}.
\end{eqnarray}
To avoid the secularity of $S_1$ over the integration of $\theta$, $\Gamma_{1t}$ is chosen to be
\begin{equation}\label{g43}
{\Gamma_{1t}} = -\left\langle F \right\rangle,
\end{equation}
where $\left \langle F \right \rangle $ means the averaging over $\theta$.
The new $\Gamma$ approximated up to $O(\varepsilon_w)$ is
\begin{equation}\label{a100}
\Gamma  = \left( {{\bf{A}}\left( {\bf{X}} \right) + \varepsilon U{\bf{b}}} \right)\cdot d{\bf{X}} + {\varepsilon ^2}\mu d\theta  - \left( {{H_0} - {\varepsilon _w}{\Gamma _{1t}}} \right)dt,
\end{equation}
with ${H_0} = \varepsilon \left( {\frac{{{U^2}}}{2} + \mu B\left( {\bf{X}} \right)} \right)$ and ${{\Gamma _{1t}}}$ given in Eq.(\ref{g43}).

By imposing the minimal action principle over the action as the integral of the Lagrangian 1-form given by Eq.(\ref{a100}) over the time, the equations of motion can be derived as
\begin{equation}\label{a115}
\mathop {\bf{X}}\limits^. {\rm{ = }}\frac{{U{{\bf{B}}^*} + {\bf{b}} \times \nabla \left( {{H_0} - {\varepsilon _w}{\Gamma _{1t}}} \right)}}{{{\bf{b}}\cdot{{\bf{B}}^*}}},
\end{equation}
\begin{equation}\label{a116}
\dot U = \frac{{ - {{\bf{B}}^*}\cdot\nabla \left( {{H_0} - {\varepsilon _w}{\Gamma _{1t}}} \right)}}{{{\varepsilon}{\bf{b}}\cdot{{\bf{B}}^*}}}.
\end{equation}

\section{Comments on the results given in Sec.(\ref{sec2})}\label{sec3}

\subsection{Violation of NIT by the coordinate transform given in Sec.(\ref{sec2})}\label{sec3.1}


Now we check that whether the coordinate transform given in Sec.(\ref{sec2}) is a NIT. In other words, whether $O(g^i)\sim O(1)$ holds for $i\in \{\mathbf{X}, \mu, U,\theta\}$. For convenience, only the pure perturbative electromagnetic potential is considered, so that the electric field only includes the inductive part and no electrostatic part exists.

To get the order sequence of $g^i$s, we first derive the order sequence of $S_1$, the equation of which is given in Eq.(\ref{g42})
with $\Gamma_{1t}$ given in Eq.(\ref{g43}). The order sequence of $S_1$ is
\begin{equation}\label{g63}
\varepsilon_w {S_1} = \varepsilon_w{\varepsilon ^{* 2}}( \cdots ) + \varepsilon_w{\varepsilon ^{*3 }}( \cdots ) +  \cdots.
\end{equation}
The superscript $*$ of ${\varepsilon ^{*2}}$  represents the order of $(\cdots)$ adjacent to it as explained before.

The lowest order term of all of $g^i$s should be of the order equaling or higher than $O(1)$ to satisfy NIT. Substituting the order sequence of $S_1$ into Eqs.(\ref{g38})-(\ref{g41}), the order sequence of $g^i$s can be derived. The lowest order of $\mathbf{g}^{X}$ is $O(1)$, which is produced by the lowest order term of the exponential expansion of the first term on the right of Eq.(\ref{g37}). The lowest order of $g^U$, $g^{\mu}$ and $g^{\theta}$ is $O(\varepsilon^{-1})$ and also originates from the lowest order term of the exponential expansion of the first term on the right of Eq.(\ref{g38}),(\ref{g39}) and (\ref{g40}), respectively. The coordinate transform for $U, \mu, \theta$ are approximately as $U_1\approx U-\varepsilon_w g_1^U, \mu_1 \approx \mu -\varepsilon_w g_1^\mu, \theta_1\approx \theta-\varepsilon_w g_1^\theta$. It's observed that for a perturbation with the amplitude being $O(\varepsilon_\omega)$, the coordinate transform amplifies the generators by $1/\varepsilon$ times to get the new coordinate. This coordinate transform doesn't satisfy NIT, as $\varepsilon$ is a very small quantity.

\subsection{The first consequence of the violation of NIT}
One consequence of the violation of NIT is as follows. In numerical and theoretical applications, the following transform between the distribution functions in the full-orbit coordinate and the gyrocenter coordinate is frequently applied
\begin{eqnarray}
f\left( {{\bf{x}},{\mu _1},{u_1},t} \right) = \smallint \begin{array}{*{20}{l}}
{F\left( {{\bf{X}},\mu ,U,t} \right)\delta \left( {{\bf{x}} - {\bf{X}} - \varepsilon \bm{\rho _0} - {\varepsilon _w}{{\bf{g}}^{{X}}}} \right)}\\
{ \times \delta \left( {\mu  - {\mu _1}} \right)\delta \left( {U - {u_1}} \right)B({\bf{X}}){d^3}{\bf{X}}d\mu dU d\theta .}
\end{array}
\end{eqnarray}
However, it's noticed that ${\varepsilon _w}g_1^U,{\varepsilon _w}g_1^\mu ,{\varepsilon _w}g_1^\theta$ given by Eqs.(\ref{g38}-\ref{g40}) are of order $O(\varepsilon_w/\varepsilon)$. So the integrand of the this integral transform should take the following formula
\begin{eqnarray}
&& \delta \left( {{{\bf{x}}} - {\bf{X}} - \varepsilon {{\bm{\rho }}_0}}-\varepsilon_w \mathbf{g}^{X} \right)\delta \left( {\mu  - {\varepsilon _w}g_1^\mu  - {\mu _1}} \right) \nonumber \\
&& \times \delta \left( {U - {\varepsilon _w}g_1^U - {u_1}} \right)\delta \left( {\theta  - {\varepsilon _w}g_1^\theta  - {\theta _1}} \right). \nonumber
\end{eqnarray}
To make modern GT self-consistent, we need to remove the violation of NIT from the coordinate transform.

\subsection{The second consequence of the violation of NIT}\label{sec3.2}

In Eqs.(\ref{a115}) and (\ref{a116}), the contribution of the perturbation are mainly $\frac{{{\varepsilon _w}{\bf{b}} \times \nabla \left\langle {U{\bf{b}}\cdot\exp \left( {\varepsilon {\rho _0}\cdot\nabla } \right){{\bf{A}}_w}\left( {{\bf{X}},t} \right)} \right\rangle }}{{{\bf{b}}\cdot{{\bf{B}}^*}}}$ and $\frac{{{\varepsilon _w}{{\bf{B}}^*}\cdot\nabla \left\langle {U{\bf{b}}\cdot\exp \left( {\varepsilon {\rho _0}\cdot\nabla } \right){{\bf{A}}_w}\left( {{\bf{X}},t} \right)} \right\rangle }}{{\varepsilon {\bf{b}}\cdot{{\bf{B}}^*}}}$, respectively. Both are not physical terms, since $\mathbf{A}_w$ includes an arbitrary gauge term like $\nabla f(x)$. The gradient operator in both terms can not cancel the gauge term.  The real physical contribution should be like $\frac{{\partial {{\bf{A}}_{\bf{w}}}/\partial t \times {\bf{b}}}}{{{\bf{b}}\cdot{B^*}}}$ and $- {\bf{b}}\cdot\frac{\partial }{{\partial t}}{{\bf{A}}_w}$, which are the $\mathbf{E}\times \mathbf{B}$ drift produced by inductive electric field, and parallel inductive electric field acceleration of  the charged particle. Therefore, Eqs.(\ref{a115},\ref{a116}) need to be modified.

Furthermore, it's found that the guiding field in Eq.(\ref{f2}) only includes the equilibrium part which is isolated out artificially from the whole magnetic field. However, for some environments in Field Reversed Pinch, Stellarator and the edge of the tokamak plasmas, it even becomes hard to distinguish the equilibrium part from the perturbative part. Therefore, for such environments, the perturbtive magnetic field should be taken into account as the guiding field.

\section{Acknowledgments}
This work is partially supported by Grants-in-Aid from JSPS (No.25287153 and 26400531) and by CSC Scholarship.

\appendix

\section{Simple introduction of Cary-Littlejohn single-parameter LTPT}\label{app1}

This theory begins with the following autonomous differential equations
\begin{equation}\label{a117}
\frac{{\partial Z_f^\mu }}{{\partial \epsilon }}\left( {{\bf{z}},\epsilon } \right) = {g_1^\mu }\left( {{{\bf{Z}}_f}\left( {{\bf{z}},\epsilon } \right)} \right),
\end{equation}
\begin{equation}\label{a118}
\frac{{d{\bf{z}}}}{{d\varepsilon }} = 0,
\end{equation}
where $\mathbf{Z}=\mathbf{Z}_f(\mathbf{z},\epsilon)$ is new coordinate, $\mathbf{z}$ is old coordinate, and $\epsilon$ is an independent variable denoting the small parameter of amplitude of perturbation.
Eqs.(\ref{a117}) and (\ref{a118}) lead to the solution
\begin{equation}\label{c1}
{\bf{z}} = \exp \left(-{{\epsilon g^i_1}{\partial _{{Z_i}}}} \right){\bf{Z}}.
\end{equation}
For a differential 1-form written as $\gamma(\bf{z})$, which doesn't depend on $\epsilon$ in the coordinate frame of $\bf{z}$, coordinate transform given by Eq.(\ref{c1}) induces a pullback transform of $\gamma$ as
\begin{equation}\label{c2}
{\Gamma _\mu }\left(\mathbf{ Z} \right) = {\left[ {\exp \left( { - \varepsilon {L_{1}}} \right)\gamma } \right]_\mu }\left( \mathbf{Z} \right) + \frac{{\partial S\left( \mathbf{Z} \right)}}{{\partial {Z^\mu }}}dZ^\mu.
\end{equation}
where $S(\mathbf{Z})$ is a gauge function and the $\mu$ component of $L_1\gamma$ is defined as ${\left( {{L_1}\gamma } \right)_\mu } = g_1^i\left( {{\partial _i}{\gamma _\mu } - {\partial _\mu }{\gamma _i}} \right)$.

When differential 1-form explicitly depends on the perturbation and can be written as
$\gamma(\mathbf{z},\varepsilon)  = {\gamma _0}(\mathbf{z}) + \epsilon {\gamma _1}(\mathbf{z}) + \epsilon ^{2}{\gamma _2}(\mathbf{z}) +  \cdots$,
Ref.(\cite{1983cary}) generalize Eq.(\ref{c2}) to be a composition of individual Lie transforms $T =  \cdots {T_3}{T_2}{T_1}$ with
\begin{equation}\label{c4}
{T_n} = \exp \left( { - \epsilon^n {L_{n}}} \right),
\end{equation}
to get the new 1-form
\begin{equation}\label{f1}
\Gamma  = T\gamma  + dS,
\end{equation}
which can be expanded by the order of $\epsilon$
\begin{equation}\label{c5}
{\Gamma _0} = {\gamma _0},
\end{equation}
\begin{equation}\label{c6}
{\Gamma _1} = d{S_1} - {L_1}{\gamma _0} + {\gamma _1},
\end{equation}
\begin{equation}\label{c7}
{\Gamma _2} = d{S_2} - {L_2}{\gamma _0} + {\gamma _2} - {L_1}{\gamma _1} + \frac{1}{2}L_1^2{\gamma _0},
\end{equation}
\begin{equation}\label{a33}
\cdots   \nonumber \\
\end{equation}
These expansion formulas can be written in a general form
\begin{equation}\label{c8}
\Gamma_n = d S_n - L_n \gamma_0 + C_n.
\end{equation}
By requiring $\Gamma_{ni}=0,i\in(1,\cdots,2N)$, the $n$th order generators are
\begin{equation}\label{c9}
g_n^j = \left( {\frac{{\partial {S_n}}}{{\partial {z^i}}} + {C_{ni}}} \right)J_0^{ij},
\end{equation}
where $J_0^{ij}$ is Poisson tensor.
And correspondingly, the $n$th order gauge function can be solved as
\begin{equation}\label{c10}
V_0^\mu \frac{{\partial {S_n}}}{{\partial {z^\mu }}} = \frac{{\partial {S_n}}}{{\partial {z^0}}} + V_0^i\frac{{\partial {S_n}}}{{\partial {z^i}}} = {\Gamma _{n0}} - {C_{n\mu }}V_0^\mu
\end{equation}
with
\begin{equation}\label{c11}
V_0^i = J_0^{ij}\left( {\frac{{\partial {\gamma _{0j}}}}{{\partial {z^0}}} - \frac{{\partial {\gamma _{00}}}}{{\partial {z^j}}}} \right)
\end{equation}
To avoid the secularity of $S_n$, usually $\Gamma_{n0}$ is chosen to be
\begin{equation}\label{c12}
{\Gamma _{n0}} = \left[\kern-0.15em\left[ {V_0^\mu {C_{n\mu }}}
 \right]\kern-0.15em\right],
\end{equation}
where $\left[\kern-0.15em\left[  \cdots
 \right]\kern-0.15em\right]$ means average over the fast variable.

%

%

%
%

\newpage
\section*{References}

\bibliographystyle{pst}
\bibliography{gyrocenter}

\end{document}